\documentclass[prl,twocolumn,showpacs]{revtex4}
\pdfoutput=1
\usepackage{amssymb,latexsym}
\usepackage{amsmath,amsbsy,bbm}
\usepackage{ifpdf}
\usepackage{epsfig,bm}
\usepackage{graphicx,comment}
\usepackage{color}
\usepackage{soul}
\usepackage{mathtools}
\usepackage[normalem]{ulem}
\begin{document}

\title{A universal neural network for learning phases and criticalities}
\author{D.-R. Tan}
\affiliation{Department of Physics, National Taiwan Normal University,
88, Sec.4, Ting-Chou Rd., Taipei 116, Taiwan}
\author{J.-H. Peng}
\affiliation{Department of Physics, National Taiwan Normal University,
88, Sec.4, Ting-Chou Rd., Taipei 116, Taiwan}
\author{Y.-H. Tseng}
\affiliation{Department of Physics, National Taiwan Normal University,
88, Sec.4, Ting-Chou Rd., Taipei 116, Taiwan}
\author{F.-J. Jiang}
\email[]{fjjiang@ntnu.edu.tw}
\affiliation{Department of Physics, National Taiwan Normal University,
88, Sec.4, Ting-Chou Rd., Taipei 116, Taiwan}

\begin{abstract}
  A universal supervised neural network (NN) relevant to compute the associated criticalities
  of real experiments studying phase transitions is constructed. The validity of the built NN
  is examined by applying it
  to calculate the criticalities of several three-dimensional (3D) models
  on the cubic lattice,
  including the classical $O(3)$ model, the 5-state ferromagnetic Potts model,
  and a dimerized quantum antiferromagnetic Heisenberg model. Particularly,
  although the considered NN is only trained one time on a one-dimensional (1D)
  lattice with 120 sites, yet it has successfully determined the
  related critical
  points of the studied 3D systems.
  Moreover, real configurations of
  states are not used in the testing stage. Instead, the employed configurations
  for
  the prediction are constructed on a 1D lattice of 120 sites and
  are based on the bulk quantities or the microscopic states of
  the considered models.
  As a result, our calculations are ultimately efficient in
  computation and the applications of the built NN is extremely broaden.
  Considering the fact that
  the investigated systems vary dramatically from each other, it is amazing that the combination of
  these two strategies in the training and the testing stages lead to a highly universal
  supervised neural network for learning phases and criticalities of 3D models.
  Based on the outcomes presented in this study, it is favorably probable that
  much simpler but yet elegant machine learning techniques can be constructed
  for fields of many-body systems other than the critical phenomena.

\end{abstract}


\maketitle

{\bf Introduction} ---
Techniques from machine learning (ML), which originally belong to the fields of
information and computer sciences, have found applications in many-body systems
during the last a few years \cite{Rup12,Sny12,Li15,Baldi:2014pta,Mnih:2015jgp,Searcy:2015apa,Baldi:2016fzo,Baldi:2016fql,Tor16,Wan16,Att16,Oht16,Hoyle:2015yha,Car16,Tro16,Bro16,Chn16,Tan16,Tubiana:2016zpw,Nie16,Mott:2017xdb,Liu16,Tam17,Xu16,Wan17,Liu17,Nag17,Kol17,Den17,Zha17,Zha17.1,Hu17,Li18,Chn18,Lu18,Bea18,Pang:2016vdc,Sha18,But18,Bar18,Zha18,Butter:2017cot,Gao18,Zha19,Gre19,Ren:2017ymm,Don19,Cavaglia:2018xjq,Dav19,Conangla:2018nnn,Li19,Lia19,Meh19,Rod19,Car19,Oht20,Larkoski:2017jix,Han:2019wue,Tan20.1,Tan20.2,Sin20,Lidiak:2020vgk,Carrasquilla:2020mas,Wan20,Son20,Cao:2020rgr,Amacker:2020bmn,Bachtis:2020dmf,Aad:2020cws,Beentjes:2020abj,Morgan:2020wvf,Shalloo:2020nhu,Tomita:2020ylz,Cabero:2020eik,Geilhufe:2020nbl,Wang:2020tgb,Nicoli:2020njz}. Such examples include studies associated
with the critical phenomena, the high energy particle physics, and the first
principles material calculations.
In some cases, the performance of
these new approaches for exploring many-body physical systems 
is comparable with that of the traditional methods.

While applying the ML techniques
to investigate physical systems has been developed for some time and many promising
achievements are obtained, research activities on improving the efficiency of those
ML methods are still vigorous. Taking topological phase transition for example,
both the supervised and the unsupervised neural networks (NN) are adopted to study this kind
of exotic criticality. Although many attempts have been done and satisfactory outcomes are reached, there is still some room for improvement.

Indeed, the accomplishments of employing the NN methods to explore the criticalities of
many-body systems are typically achieved using very complicated infrastructures such
as the convolutional neural networks (CNN). Specially, a convolutional layer,
which is intended to capture certain characteristics of the considered
system, is included in the CNN. From this point of view, an investigation
on the targeted system somehow is required before one can build a
appropriate CNN to study the associated criticality. In other words,
a constructed CNN for a particular model may not be applicable for another one.
In particular, redesigning a relevant convolutional layer is needed whenever a
new and completely different system is considered. 

Apart from the complications introduced in the previous paragraph,
in the training stage of a NN calculation, few to several
thousand real configurations generated by
some (numerical) methods are used as the training set typically.
Moreover, each of the system sizes used in the calculations
requires one separate training as well.
This described training procedure is time consuming and may use up a lot
of computing resources. Besides, the majority of NN calculations
for studying critical phenomena also employ real configurations of states
in the testing (prediction) stage. Such a strategy in the testing stage
take huge amount of storage space if system(s) of large sizes are considered, and
it may not directly applicable to some results obtained from experiments.
This is because in many cases it is
the bulk properties, not the detailed states of the studied
materials, is measured and recorded in the laboratories. As a result, it will be highly desirable to
construct a NN with an infrastructure that can simplify the training procedure
and make a closer connection to the reality conducted in experiments.

As demonstrated in Refs.~\cite{Li18,Tan20.1}, when studying the phase transitions of the $q$-state ferromagnetic Potts
models with a NN, the consideration of using the theoretical ground
state configurations in the ordered phase as the training sets greatly reduces the needed
computations for the training. Later a training strategy of using only two configurations made artificially
can even detect the critical points with high accuracy for the three-dimensional (3D)
classical $O(3)$ models, the two-dimensional (2D) $XY$ models, and
the 3D and 2D dimerized quantum antiferromagnetic Heisenberg models \cite{Tan20.2}.
The outcomes shown in Ref.~\cite{Tan20.2} indicate that little information of the underlying systems
is sufficient to uncover the critical phenomena of these considered models.
This is quite surprising, considering the fact that these studied models have infinite
number degree of freedom. Moreover, it is amazing as well that such a simple training
procedure is valid for models that are different dramatically from each other.

Despite the remarkable achievements, one training is still needed for every considered system size
for the NN calculations carried out in Ref.~\cite{Tan20.2}. In addition, real configurations
of states of the studied models are stored and employed in the testing stage as well. As pointed out previously,
this does not match very well with many real experimental situations, since
typically
the bulk quantities rather than the detailed microscopic details are available in experiments.
Here we demonstrate that it is possible to build a simple supervised NN that is relevant to
compute the associated criticalities of real experiments studying phase transitions.
Moreover, the NN is obtained by carrying out only one training on a fixed size one-dimensional (1D) lattice,
and the validity of the resulting
NN is examined by applying it to calculate the critical points of various models with many system sizes.
Specifically, we train a MLP (defined later) on a 1D lattice of 120 sites and
successfully employ the obtained weight to calculate the critical points
of the 3D classical $O(3)$ model, a 3D dimerized quantum antiferromagnetic Heisenberg model (abbreviated as the plaquette model here),
and the 3D 5-state ferromagnetic Potts model. Moreover, 
instead of using the configurations of states directly determined from the simulations,
in the prediction stage, the configurations employed are constructed based on
some bulk quantities of the considered models having certain characteristics
(These characteristics will be detailed in the relevant sections).
Such a strategy for the testing stage is not only economically efficient in storage, 
but is also close to the situations of real experiments, namely only the bulk quantities
are recorded. As a result, the procedures of the NN calculations presented here are
directly applicable for the relevant experiments.

Considering the fact that the majority of applying ML methods to study many-body systems use real microscopic configurations for the prediction stage, here we
also construct configurations on
a 1D lattice of 120 sites based on the microscopic states of the studied systems. It turns out that this
second method of building the configurations for the NN prediction leads to successful outcomes as well.

It is beyond one's intuition and is surprisingly impressive that a NN determined
on a fixed 1D lattice can be used to study the criticalities of several 3D
models of various system sizes. In particular, these studied 3D systems
are dramatically different among themselves. Finally, in conjunction with the data obtained from
the relevant experiments, our approach is perfectly suitable for studying the associated
phases and criticalities.

{\bf The constructed supervised Neural Network} ---
In this section, we review the supervised NN, namely the multilayer perceptron
(MLP) used in our study. The employed training set and label will be introduced as well.
Part of this section is a short summary of that in Refs.~\cite{Tan20.1,Tan20.2}.

The MLP used in our investigation is already detailed in Ref.~\cite{Tan20.1,Tan20.2}. 
Specifically, using the NN library keras and tensorflow \cite{kera,tens},
we construct a supervised NN
which consists of only one input layer, one hidden layer of 512 
independent nodes, and one output layer. In addition, The algorithm, optimizer, and
loss function considered in our calculations are the minibatch, the adam, and the
categorical cross entropy, respectively.
To avoid overfitting, we also apply $L_2$ regularization at various stages.
The activation functions employed here are ReLU and softmax.
The details of the constructed MLP, including the steps of
one-hot encoding and flatten (and how these two processes work)
are shown in the supplemental materials and are available in Refs.~\cite{Tan20.1,Tan20.2}.

Finally, for the three studied models, results calculated using 10 sets of random seeds
are all taken into account when presenting the final outcomes. Specifically,
each set of random seeds leads to a mean resulting from a few hundred to several thousand (independent)
outcomes. The final quote values
are then based on the mean and the standard deviation of these 10 mean results.

For the training set and the associated training procedure carried
out in our NN calculations,
a simple alternative is considered. Specifically, instead of using
real configurations obtained from the simulations or
the theoretical ground states in the ordered phase of
the studied system(s), only one training
on a 120 sites one-dimensional (1D) lattice is conducted.
In addition, the training set consists of two configurations.
Particularly, one of the configuration has 0 as the value
for each of its site, and every spot of the other configuration
takes the number of 1.
Consequently, the output labels are naturally the vectors of $(1,0)$ and $(0,1)$.

In this study, instead of using real configuration of states for the testing (prediction), bulk quantities are
considered. Based on the fact that physical observables which take distinct values at different phases
is essential to distinguish various states, 
we use a criterion to choose the relevant bulk quantities to construct configurations for the testing (prediction).
Specifically, in the simulated range of $T$ (or other relevant parameters) the chosen observable(s) should saturate
to distinct values at the high and the low temperature regions. We find that the Binder ratios $Q_1$ and $Q_2$ suit
this criterion very well.
After a desired observable $O$ is picked, one carries out the associated simulations with many different
inverse temperatures $\beta$ and
on lattices of various linear sizes $L$.
For each $L$ the difference of $O$ at the highest and the lowest temperatures is recorded.
Let this quantity be denoted by $D_{O,L}$. Then for the same observable at every other temperature,
a configuration on a 1D lattice of 120 sites, which will be used to feed the trained NN,
is constructed through the following steps. 
\begin{enumerate}
  \item{For a given temperature $T_{m}$, let the difference between the values of $O$ at $T_{m}$ and that at the lowest temperature
    be $D_{m}$.}
  \item{For each site $i$ of the 1D lattice, choose a number $p$ in [0,1) randomly and uniformly.}
  \item{If $p \ge |D_{m}/D_{O,L}|$, then the site $i$ is assigned the integer $1$. Otherwise the integer $0$ is given to $i$. }
   \item{Repeat above steps for all the simulated (inverse) temperatures.}
  \end{enumerate}

The configurations constructed from these described steps, which are based on the bulk quantities of the studied models,
are then fed to the trained NN. With this set up, the associated critical point can be estimated by investigating the
temperature (or other relevant tuning parameter) dependence of the output vectors $\vec{V}$. In particular, when those
$\vec{V}$ are considered as functions of $\beta$, the critical point
is located at the inverse temperature corresponding to the intersections of the two curves made up from the components of $\vec{V}$.
It should be pointed out that, the critical point may be also obtained as the (inverse) temperature associated with the output which
has the smallest value of magnitude.

We would like to emphasize the fact that technically speaking, only one training on a pre-chosen 1D lattice
using two objects as the training set is conducted here.
The resulting weight is then used to perform the NN prediction for all the studied 3D models
with various system sizes. Hence the training process
takes much less computing resources than (any) other known schemes in the literature. Moreover, no real configuration
of states are used in the testing stage, and only bulk quantities fulfilling certain properties are recorded. As a result,
the required storage is also much smaller when compared with other methods of studying phases and criticalities using the
NN approaches. 

Despite its simplicity in the training as well as the testing stages, as we will demonstrate later, the trained NN
is capable of detecting the critical points of all the considered models with high accuracy.

Apart from the constructed 1D configurations introduced above, based on the microscopic states of the studied systems,
configurations on a 1D lattice of 120 sites
are built additionally and are used for the NN prediction as well. This second method for the testing stage leads
to successful calculations as well. The details of this second construction will be presented in the supplemental materials.

{\bf The numerical results} ---
In this section, results obtained by using the constructed NN will be presented. The needed Monte Carlo
simulations are performed using the Wolff algorithm \cite{Wol89} and the stochastic series expansion algorithm with
operator-loop update \cite{San99}.

As mentioned earlier, for all the three studied 3D models, we find the first and the second Binder ratios ($Q_1$ and $Q_2$)
suit the criterion introduced previously. After the appropriate observables are determined,
two types of configurations used for the testing (prediction) are constructed according
to the rules outlined in the previous section and the supplemental materials.

\begin{figure}
\vskip-0.5cm
\begin{center}

\includegraphics[width=0.4\textwidth]{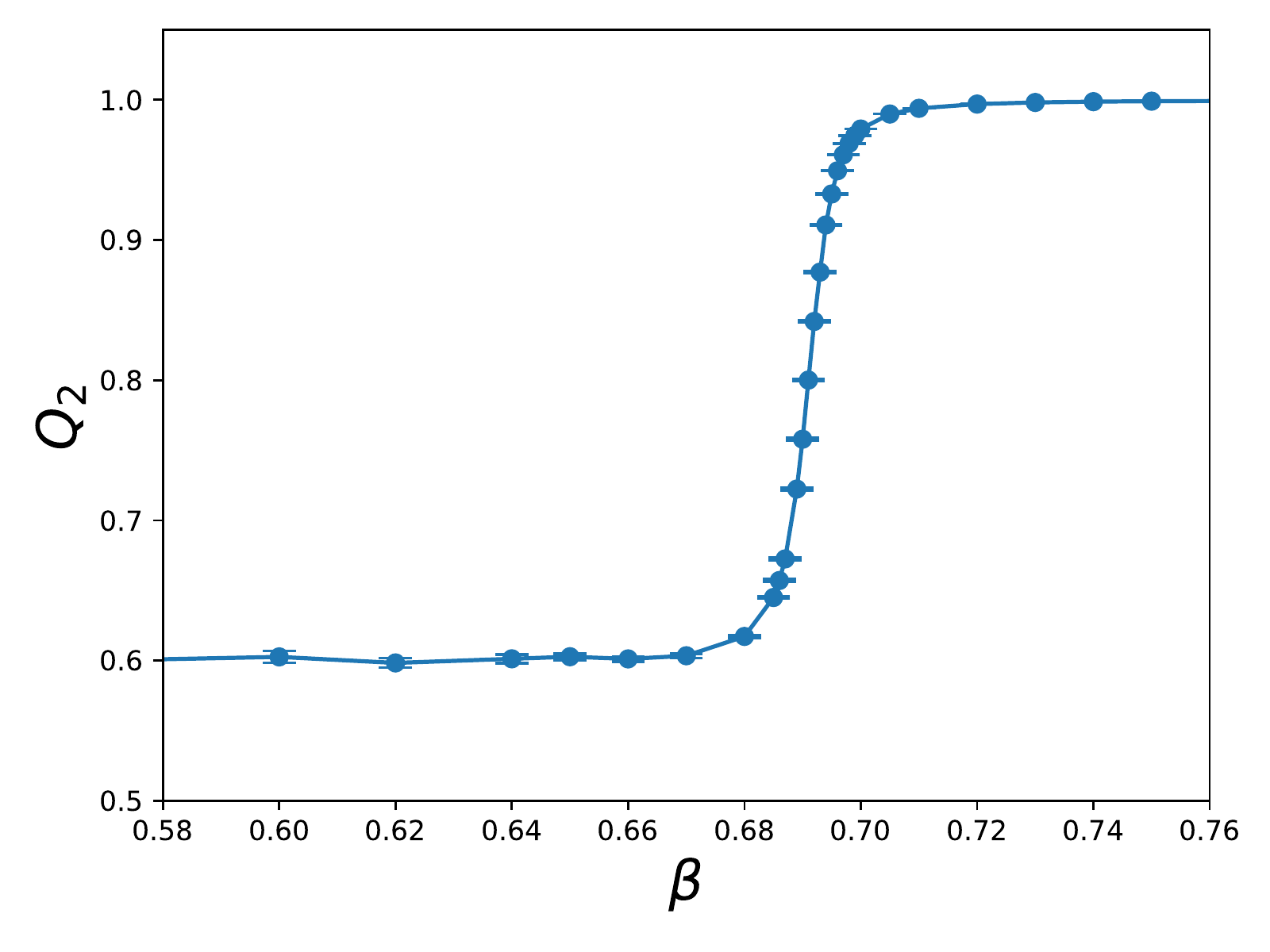}

\end{center}\vskip-0.7cm
\caption{$Q_2$ (on $48^3$ cubic lattices)
  as a function of $\beta$ for the 3D classical $O(3)$ model.}
\label{Q1Q2_O3}
\end{figure}

The observable $Q_2$ (obtained on a $48^3$ cubic lattice) as a function of $\beta$ for the 3D classical $O(3)$ model is shown in fig.~\ref{Q1Q2_O3}. 
Although the outcomes in fig.~\ref{Q1Q2_O3} indicate that $T_c$ lies between $\beta = 0.67$ and $\beta = 0.71$,
it can not determined directly from the results in the figure.

\begin{figure}
\vskip-0.5cm
\begin{center}

\includegraphics[width=0.4\textwidth]{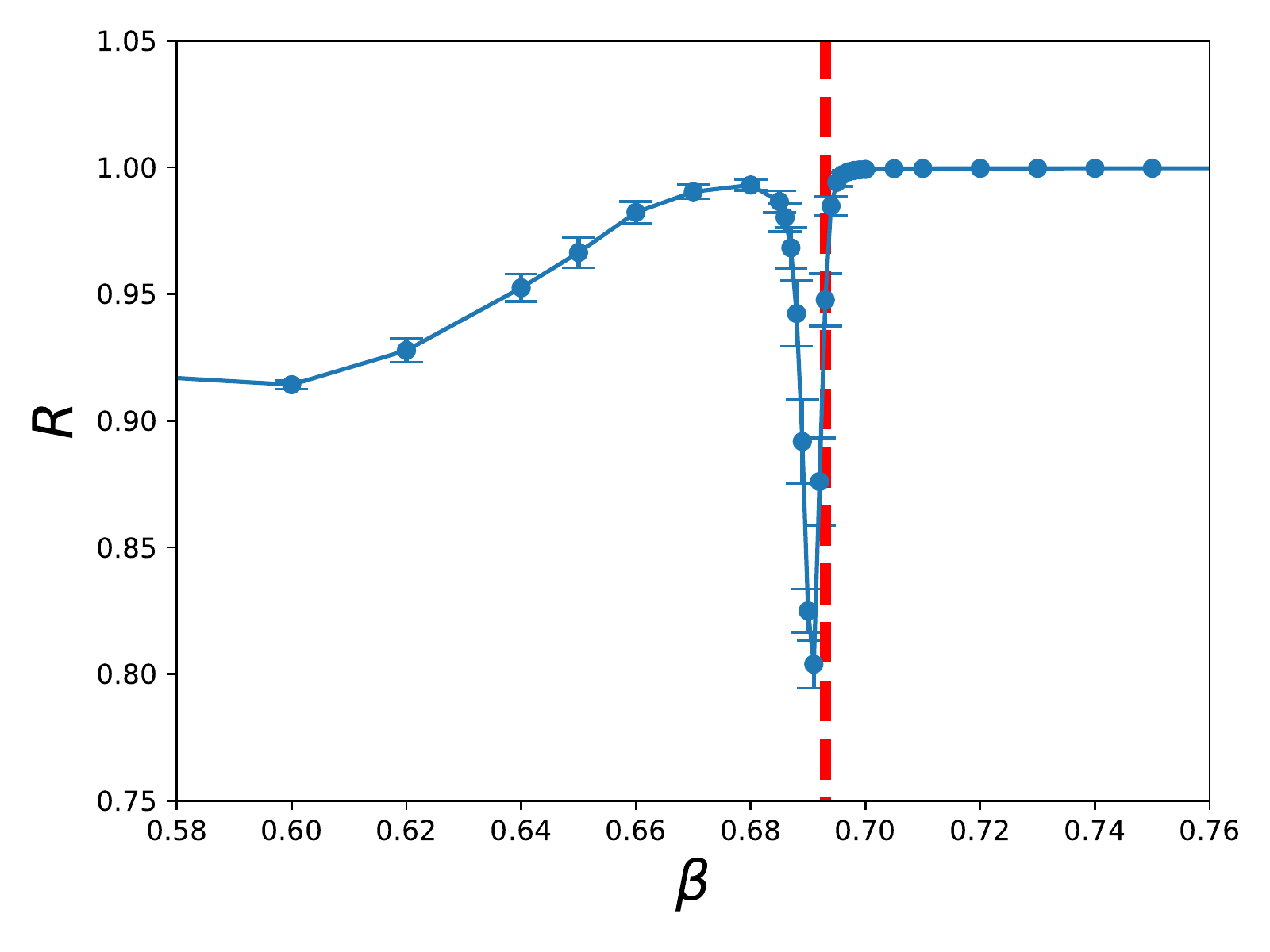}

\end{center}\vskip-0.7cm
\caption{$R$
  as a function of $\beta$ for the 3D classical $O(3)$ model. The dashed line is the expected $\beta_c$.
The results are associated with $Q_2$ (on $48^3$ cubic lattices).}
\label{RCQ1Q2_O3}
\end{figure}

\begin{figure}
\vskip-0.5cm
\begin{center}

  \includegraphics[width=0.4\textwidth]{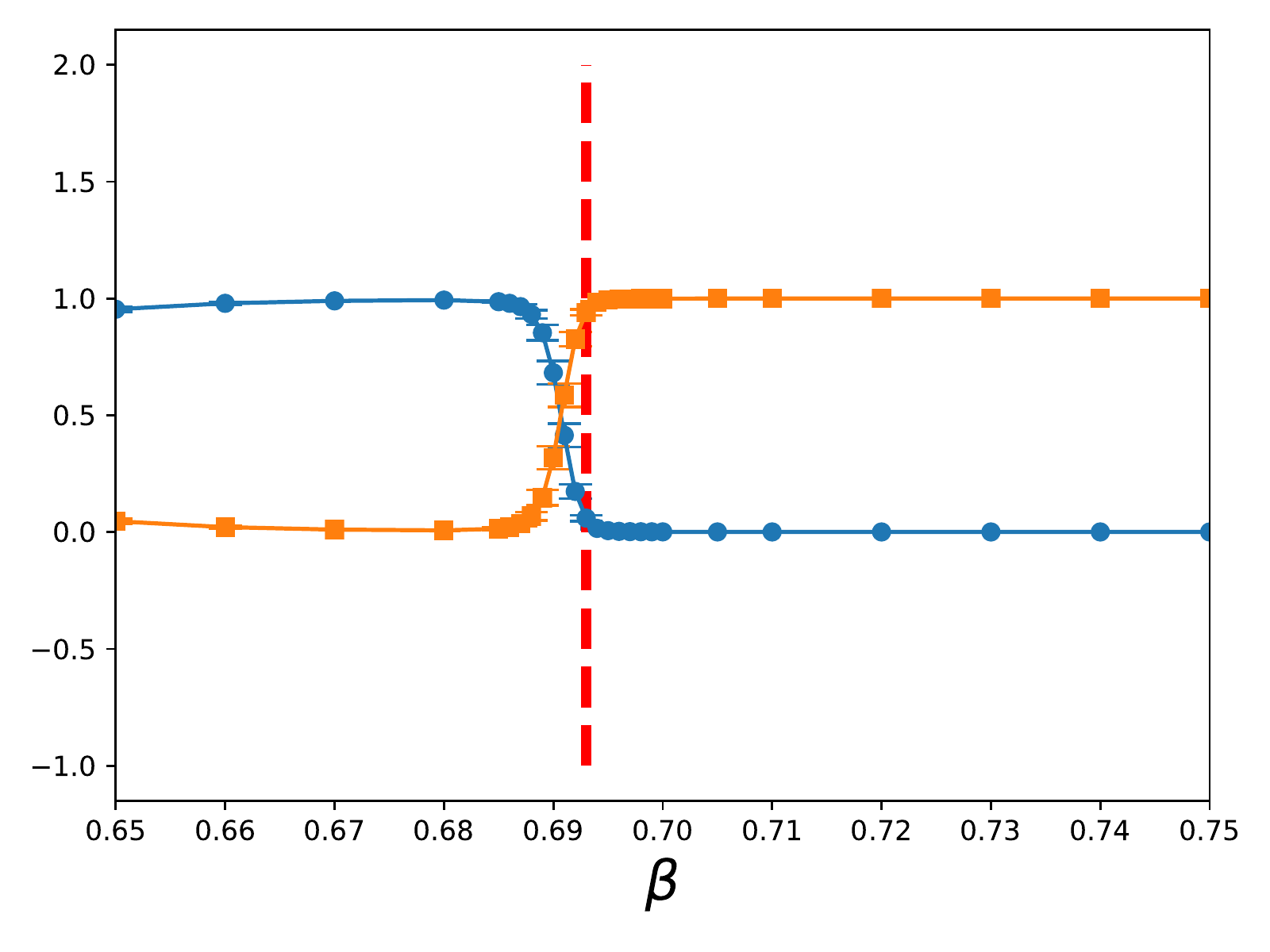}

\end{center}\vskip-0.7cm
\caption{Individual component of the output vectors 
  as functions of $\beta$ for the 3D classical $O(3)$ model.
  The results are related to the observable $Q_2$ (on $48^3$ cubic lattices).
The dashed line is the expected $\beta_c$.}
\label{components_Q1Q2_O3}
\end{figure}

The norm $R$ (associated with $Q_2$ obtained on a $48^3$ lattice) of the outputs, which are determined by using the NN weight obtained from the
calculation of employing only two training objects on a 1D
lattice of $L=120$, are shown in fig.~\ref{RCQ1Q2_O3}. As can be seen from the figure, the minimum of $R$ takes place at a $\beta$ very close to the critical inverse temperature $\beta_c$ (dashed vertical lines) of the
considered 3D classical $O(3)$ model.
Moreover, the $\beta$ dependence of individual component $V_1$ and $V_2$ of the output vectors $\vec{V}$ are demonstrated in fig.~\ref{components_Q1Q2_O3}. Similar to the scenario of $R$, the intersection of the $V_{1,2}$ v.s. $\beta$ curves is very close
to $\beta_c$ as well. The results in fig.~\ref{RCQ1Q2_O3} and \ref{components_Q1Q2_O3}
confirm the validity of our NN approach outlined in the previous sections.

Similar scenarios are observed for the 3D 5-state ferromagnetic Potts model and the 3D dimerized
quantum plaquette model, see the supplemental materials.

\begin{figure}
\vskip-0.5cm
\begin{center}

\includegraphics[width=0.4\textwidth]{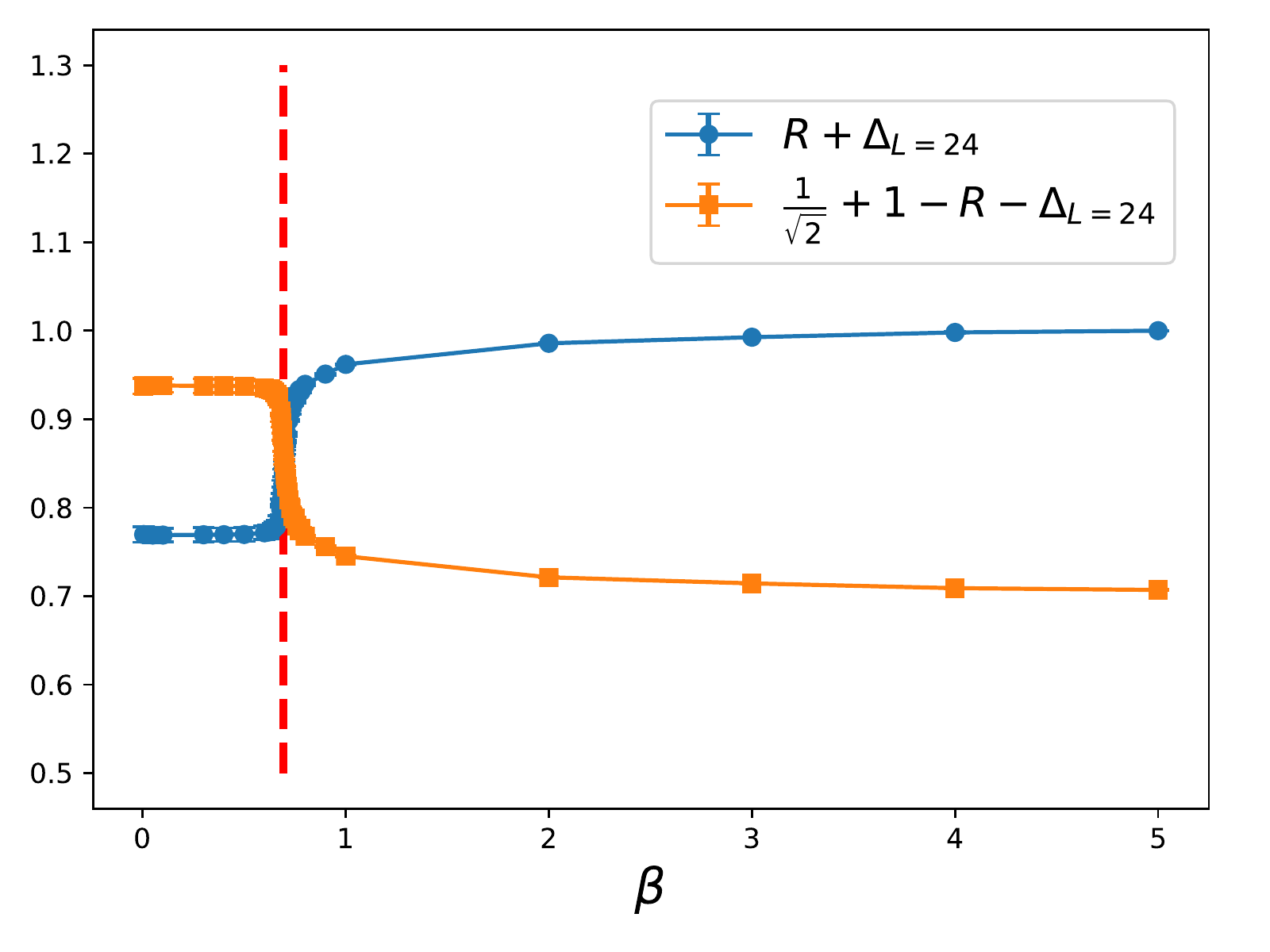}
\end{center}\vskip-0.7cm
\caption{$R+\Delta$ and $1/\sqrt{2}+1-R-\Delta$ as functions of $\beta$ for the 3D classical $O(3)$ model.
The dashed line is the expected $\beta_c$. The outcomes are associated with the spin configurations on $24^3$ lattices. The definitions of $R$ and $\Delta$ can be found in Ref.~\cite{Tan20.2}.}
\label{CONFO3}
\end{figure}

Finally, configurations on a 1D lattice of 120 sites using the microscopic
classical $O(3)$ spin states obtained from the simulations
are built and are used for the prediction. We first randomly and uniformly pick up 120 sites of the spin states
and then follow the procedures outlined in Ref.~\cite{Tan20.2} to construct the associated configurations for the prediction.
It turns out that this second approach leads to high precision determination of the critical point(s) as well, see
fig.~\ref{CONFO3} and the supplemental materials.

{\bf Discussions and Conclusions} ---
In this study, we calculate the critical points associated with the 3D classical $O(3)$ model, the 3D 5-state ferromagnetic Potts model,
and a 3D dimerized quantum antiferromagnetic Heisenberg model, using the technique of supervised Neural Networks.
Particularly, unlike any NN studies known in the literature, an extremely simple built NN as well as elegant
training and testing procedures are adopted in our investigation.

In our approach for the training stage, only one training using two configurations on a fixed size (120) 1D lattice
is conducted. The resulting weight is then applied to calculate the critical points of three 3D systems. In the testing (prediction)
stage, not the real configurations, but the configurations constructed based on the bulk observables which have certain
characteristics are employed to feed the trained NN. The combination of such two strategies for the training and the testing stages
are not only highly efficient in computation, but also determine accurately the critical points of the considered 3D models.
Apart from this extremely efficient approach, we also construct configurations on a 1D lattice of 120 sites based on the
microscopic spin states for the prediction. The second method leads to successful determination of the targeted critical point(s) as well.

Although the nature and the degree of freedom of the three considered models vary dramatically from
each other, it is remarkable that the constructed NN is applicable for all of these cases. Specially,
it is amazing the weight obtained by carrying out only one training procedure on a 1D lattice with fixed size
can be employed successfully to calculate the critical points associated with various 3D models. Here we would like
to emphasize the fact that it seems flexible to choose a lattice of any dimension and size for the training. Indeed, as we will shown in the
supplemental materials, the targeted critical points can be determined precisely using a NN trained on a 1D lattice of 200 sites as
well as a 2D square lattice with linear system size $L=16$. 

When compared with other NN schemes, our approach is highly advanced in both the efficiency and the simplicity. Some benchmarks are provided in the 
supplemental materials.
In reality, bulk quantities are measured in many experiments. Hence the use of employing
bulk observables to build the associated configurations to feed the trained NN makes the approach presented
here more attractive since it is directly applicable to the data obtained from the relevant experiments. 

Finally, according to our series of studies regarding improving the efficiency of applying the
supervised NN to investigate the phase transitions \cite{Li18,Tan20.1,Tan20.2}, it is probable
that with some elegant ideas, much more efficient machine learning methods for studying many-body physical problems other than
the critical phenomena can be obtained.

Partial support from Ministry of Science and Technology of Taiwan is 
acknowledged.

\vskip0.5cm
{\bf Supplemental materials} 

For the benefit of readers, in this supplemental materials, the constructed NN, the studied models, and more results are presented.
We would like to emphasize the fact that part of this supplemental materials are available in Refs.~\cite{Li18,Tan20.1,Tan20.2}.

\begin{figure*}
  \begin{center}
      \includegraphics[width=0.8\textwidth]{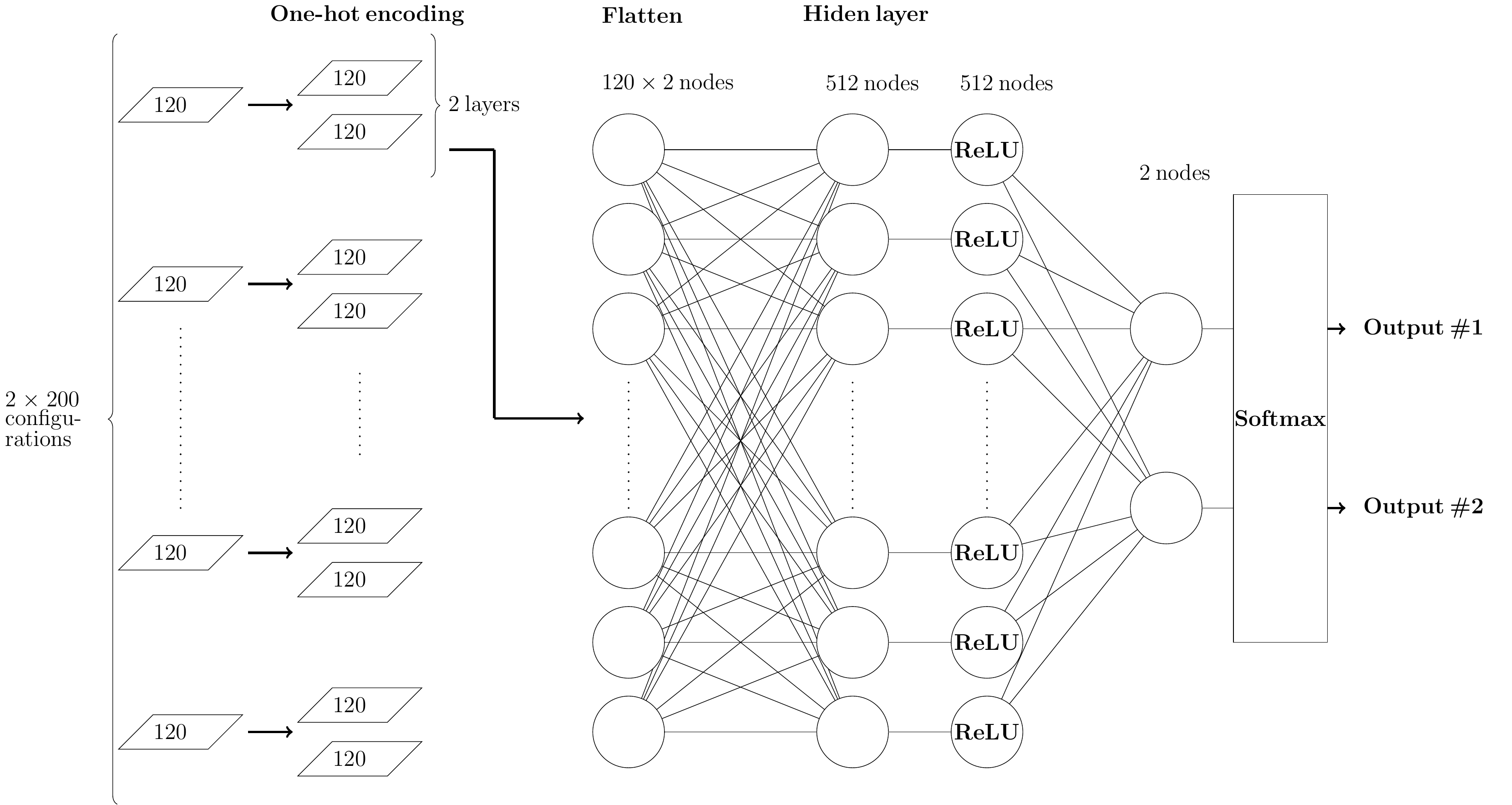}
\end{center}\vskip-0.7cm
\caption{The NN (MLP), which consists of one input layer, one hidden layer,
  and one output layer, used here and in Refs.~\cite{Tan20.1,Tan20.2}.
  The objects in the input layer are
  made up of 200 copies of only two configurations.
  There are 512 nodes in the hidden layer and each of these
  nodes is independently connected to every object in the input layer. Before
  each training object is connected to the nodes in the hidden layer,
  the steps of one-hot encoding and flatten are applied. The activation
  functions (ReLU and softmax) and where they are employed are demonstrated
  explicitly. Finally, the output layers consist of two elements.}  
\label{MLP}
\end{figure*}

\vskip0.5cm
{\bf The constructed NN} ---
The built NN mentioned in the main text consists of only one input layer, one hidden layer of 512 nodes, and one output layer.
Figure \ref{MLP} represents the infrastructure of the NN (MLP) employed in this investigation. For the reader who
are interested in the details of the considered NN (MLP) shown in fig.~\ref{MLP}, including the used algorithms, the optimization,
the activation functions ReLU and softmax, and the loss function categorical crossing entropy, are referred to Ref.~\cite{Tan20.2}.

\vskip0.5cm
{\bf The microscopic models and observables} ---
The models considered in this study have been investigated extensively in the
literature. Here we summarize the
associated Hamiltonians and the observables that are relevant to our study.
Some of these descriptions have been appeared in Refs.~\cite{Li18,Tan20.1,Tan20.2}.
\vskip0.25cm
--- The 3D classical $O(3)$ (Heisenberg) model \cite{Tan20.2}---

The Hamiltonian $H_{O(3)}$ of the 3D classical $O(3)$ (Heisenberg) model on a cubic lattice
considered in our study is given by 
\begin{equation}
\beta H_{O(3)} = -\beta \sum_{\left< ij\right>} \vec{s}_i\cdot\vec{s}_j,
\label{eqn1}
\end{equation}
where $\beta$ is the inverse temperature and $\left< ij \right>$ stands for 
the nearest neighbor sites $i$ and $j$. In addition, in Eq.~(\ref{eqn1})
$\vec{s}_i$ is a unit vector belonging to a 3D sphere $S^3$ and is located at
site $i$. 

Relevant observables used in this study are the first and the second Binder
ratios ($Q_1$ and $Q_2$) which are given as
\begin{eqnarray}
  Q_1 &=& \langle |m| \rangle^2/ \langle m^2\rangle, \\
  Q_2 &=& \langle m^2\rangle^2 / \langle m^4 \rangle,
  \end{eqnarray}
where $m = \frac{1}{L^3}\sum_i \vec{s}_i$ and $L$ is the linear box size of the system \cite{Bin81}. 
\vskip0.25cm
--- The 3D dimerized quantum antiferromagnetic Heisenberg models \cite{Tan20.2}---

\begin{figure}
  \begin{center}
\includegraphics[width=0.3\textwidth]{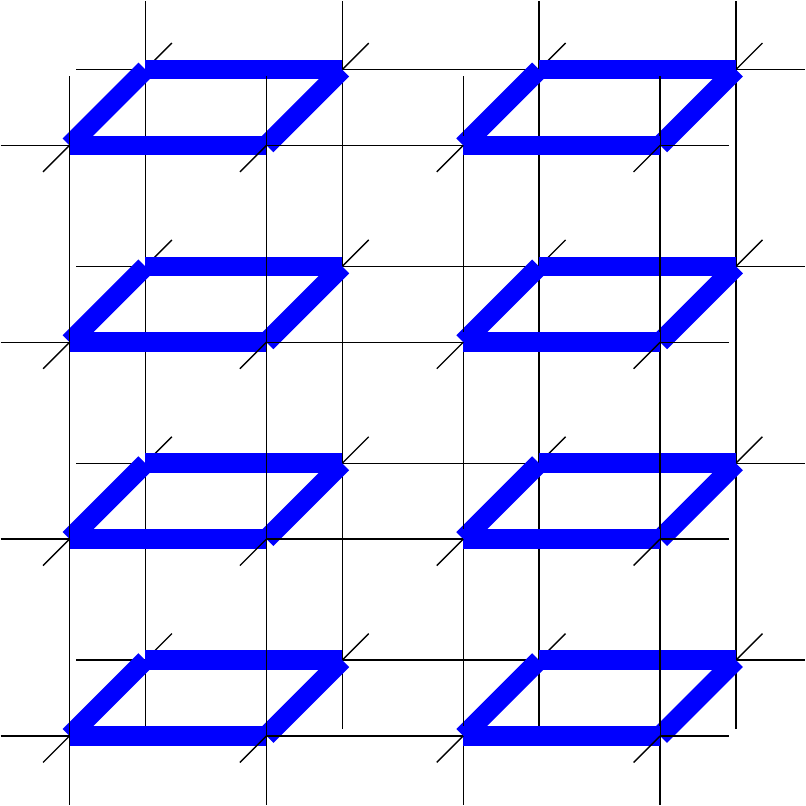}
  \end{center}\vskip-0.7cm
  \caption{The 3D dimerized plaquette quantum antiferromagnetic
    Heisenberg model studied in this investigation \cite{Tan20.2}.}
\label{model_quantum}
\end{figure}

The 3D dimerized quantum antiferromagnetic Heisenberg model has the following
expression as its Hamiltonian

\begin{eqnarray}
  H = \sum_{\langle i, j \rangle}J_{ij} \vec{S}_i \cdot \vec{S_j},
  \label{eqn2}
\end{eqnarray}
where again $\langle i, j\rangle$ stands for the nearest neighbor sites $i$ and $j$,
$J_{ij} > 0$ is the associated antiferromagnetic coupling (bond) connecting $i$ and
$j$, and $\vec{S}_i$ is the spin-1/2 operator located at $i$.
The cartoon representation of the studied model is shown in fig.~\ref{model_quantum}
and this model will be named 3D plaquette model here if no confusion
arises. Moreover, in the figure, the antiferromagnetic couplings for
the thick and thin bonds are $J'$ and $J$, respectively.
Based on fig.~\ref{model_quantum}, one sees that as the magnitude of $g$
(Which is defined as $J'/J$) increases, a quantum phase transition from
the ordered to the disordered states will occur once $g$
exceeds a particular value $g_c$. Relevant observables
considered in our investigation for
studying the quantum phase transition are again the first and the second Binder ratios $Q_1$
and $Q_2$. For the studied spin-1/2 system, $Q_1$ and $Q_2$ have the following definitions

\begin{eqnarray}
  Q_1 &=& \langle |M_s| \rangle^2/ \langle M_s^2\rangle, \\
  Q_2 &=& \langle M_s^2\rangle^2 / \langle M_s^4 \rangle,\\
  M_s &=& \frac{1}{L^3} \sum_{i} (-1)^{i_1 + i_2} S_i^z. 
  \end{eqnarray}

\vskip0.25cm
--- The 3D 5-state ferromagnetic Potts model \cite{Tan20.1}---

The Hamiltonian $H_{\text{Potts}}$ of the 5-state Potts model on the 3D cubic
lattice considered in our study is given by \cite{Wu82,Swe87,Wan89,Wan90}

\begin{equation}
\beta H_{\text{Potts}} = -J\beta \sum_{\left< ij\right>} \delta_{\sigma_i,\sigma_j},
\label{eqn3}
\end{equation}
where $J > 0$, $\beta$ is the inverse temperature and similar to the
convention introduced earlier, here $\left< ij \right>$ stands for
the nearest neighbor sites $i$ and $j$. In addition, in Eq.~(\ref{eqn3})
the $\delta$ refers to
the Kronecker function and finally, the Potts variable
$\sigma_i$ appearing above at each site $i$ takes an integer value from
$\{1,2,3,4,5\}$.

The observables considered here for the 3D 5-state ferromagnetic Potts model
are again the first and the second Binder ratios ($Q_1$ and $Q_2$) which
are expressed as
\begin{eqnarray}
  Q_1 &=& \langle |m| \rangle^2/ \langle m^2\rangle, \\
  Q_2 &=& \langle m^2\rangle^2 / \langle m^4 \rangle,
  \end{eqnarray}
where on a cubic lattice with a linear box size $L$ the $m$ appearing above
is defined as
\begin{equation}
m = \frac{1}{L^3} \sum_{j} \exp\left(i\frac{2\pi \sigma_j}{5}\right),
\end{equation}
and the summation is over all sites.

As the temperature $T$ changes from low to high,
phase transitions will take place for the 3D classical $O(3)$
and the 5-state ferromagnetic Potts models.
The critical points $T_c$ of the 3D classical $O(3)$ and the 5-state
ferromagnetic Potts models
as well as the $g_c$ of the 3D plaquette model described above have been
calculated with high accuracy in the literature \cite{Hol92,Cam02,Has05,San10,Tan18}.

\begin{figure}
  \begin{center}
    \vbox{
\includegraphics[width=0.4\textwidth]{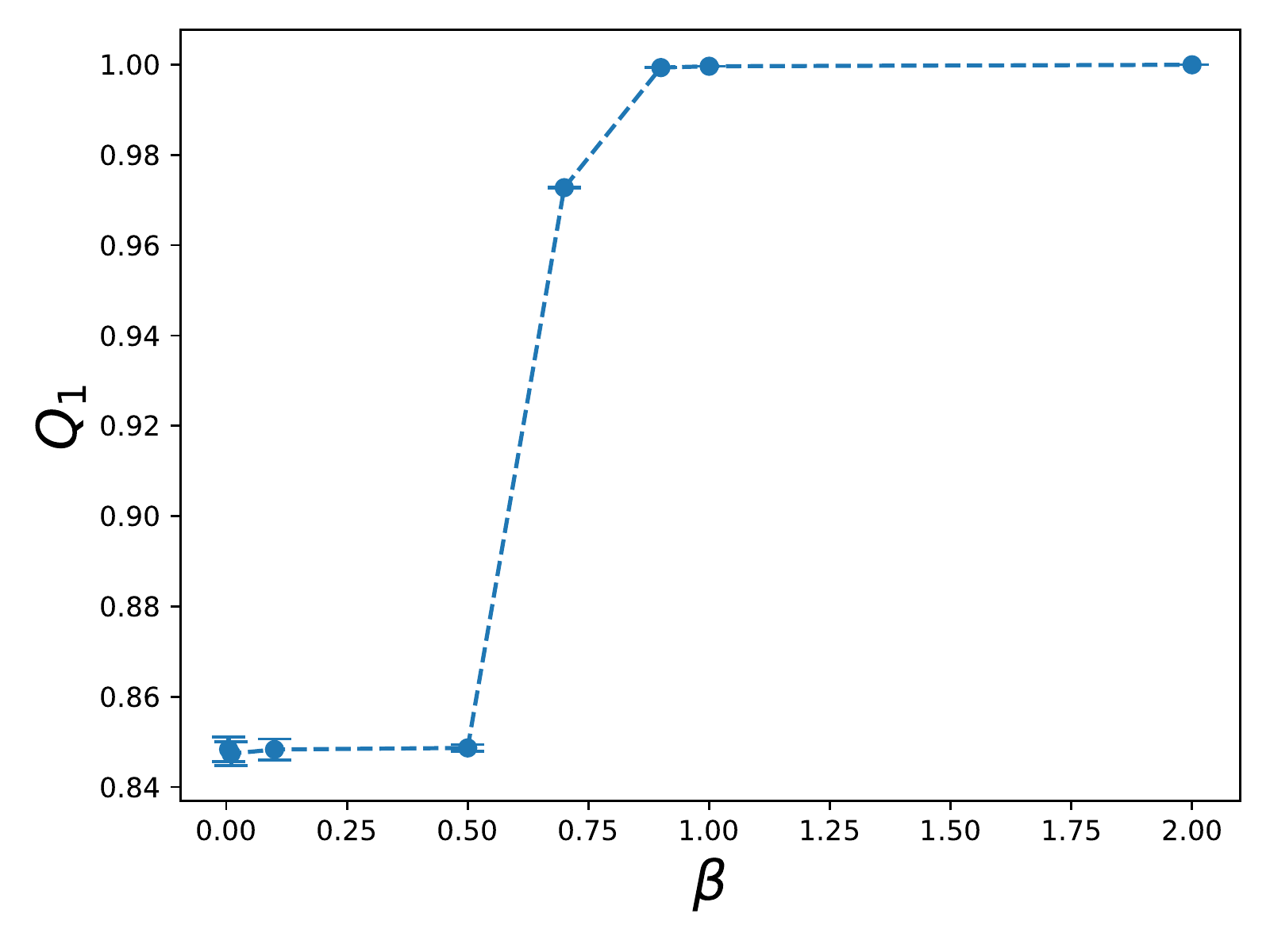}
\includegraphics[width=0.4\textwidth]{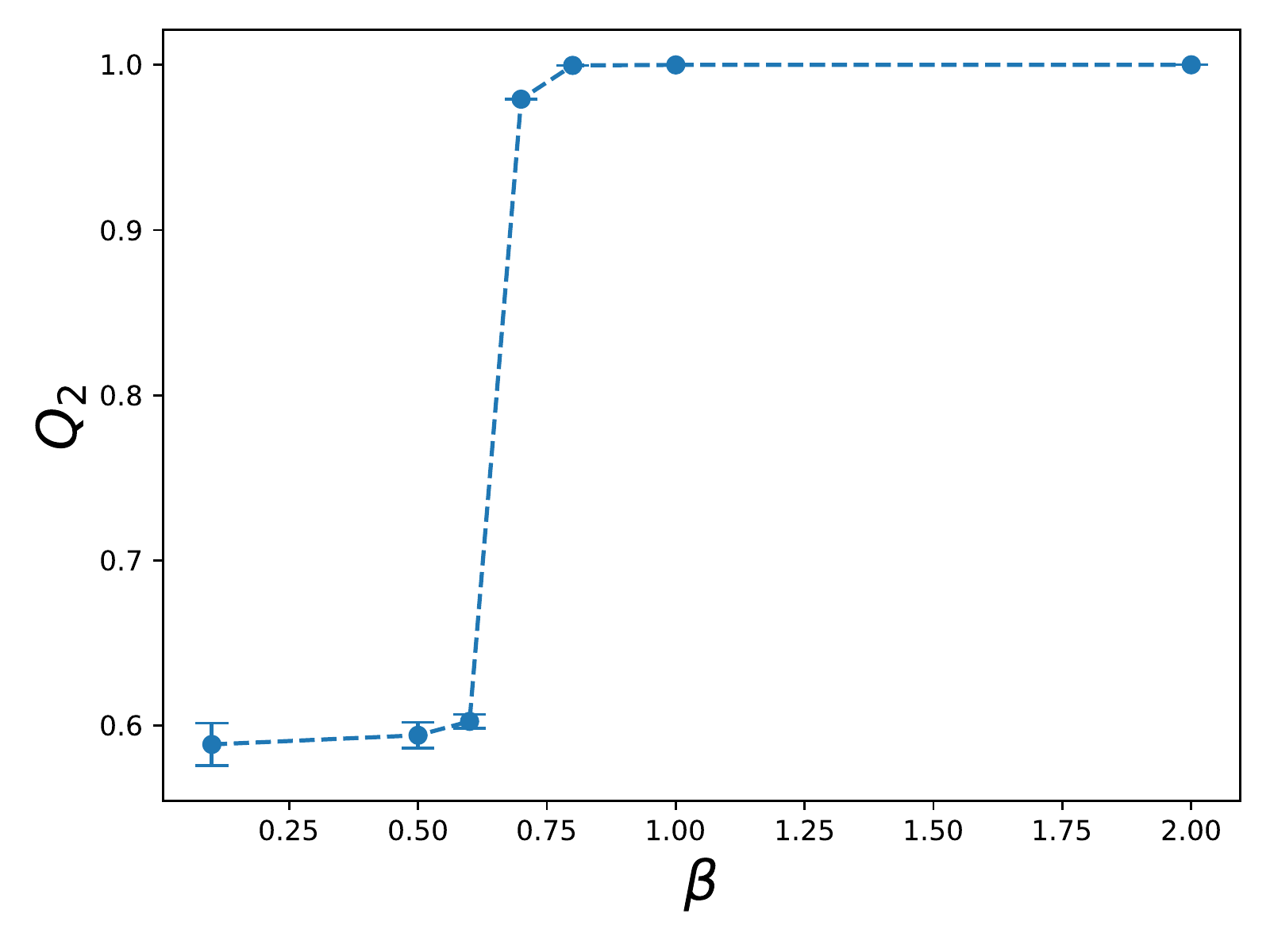}
    }
  \end{center}\vskip-0.7cm
  \caption{$Q_1$ (top panel, obtained on $12^3$ lattices) and $Q_2$ (bottom panel, obtained on $48^3$ lattices)
    as functions of $\beta$ for the 3D classical $O(3)$ model.}
\label{O3_MC}
\end{figure}

\begin{figure}
  \begin{center}
    \vbox{
\includegraphics[width=0.4\textwidth]{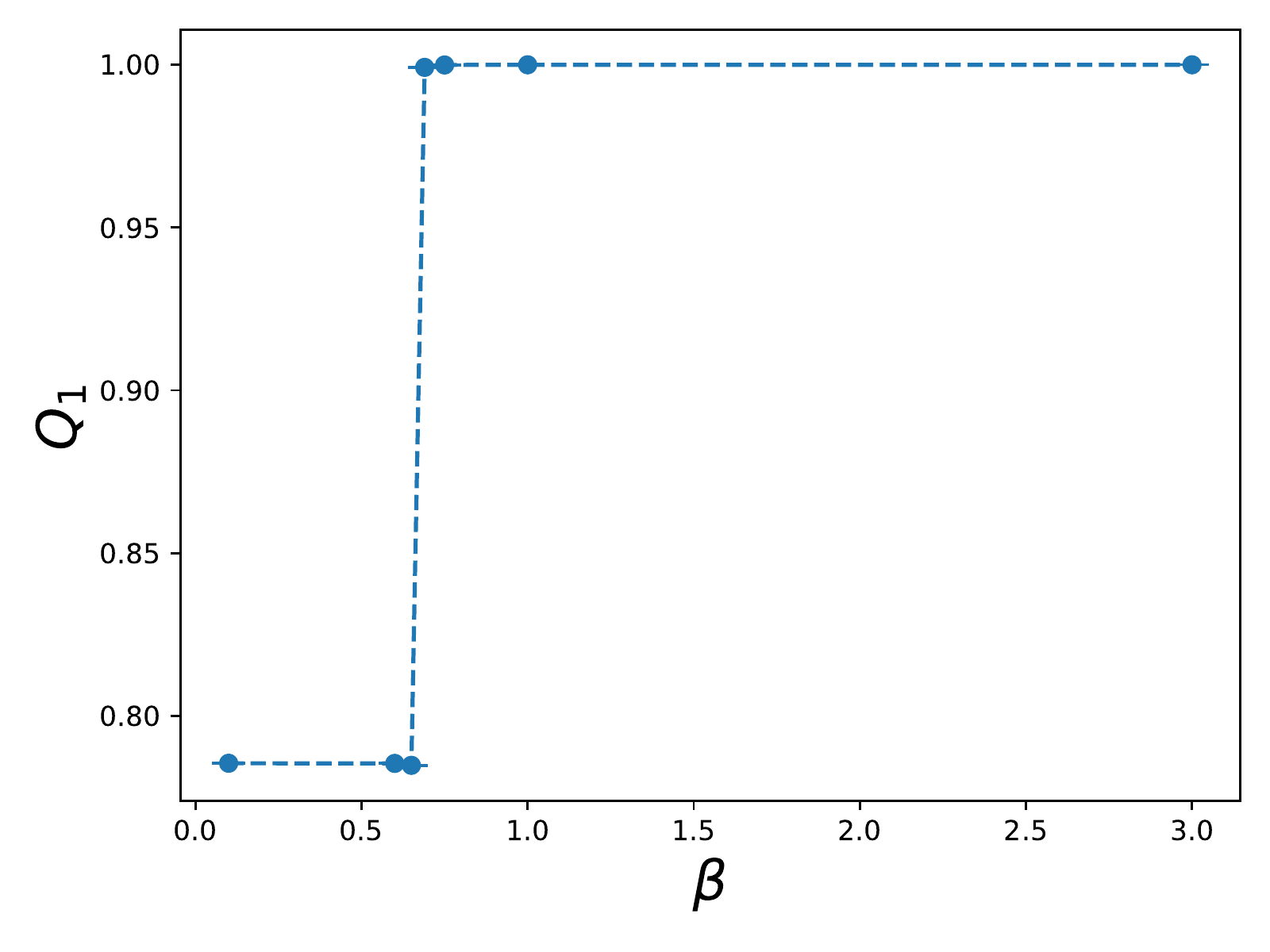}
\includegraphics[width=0.4\textwidth]{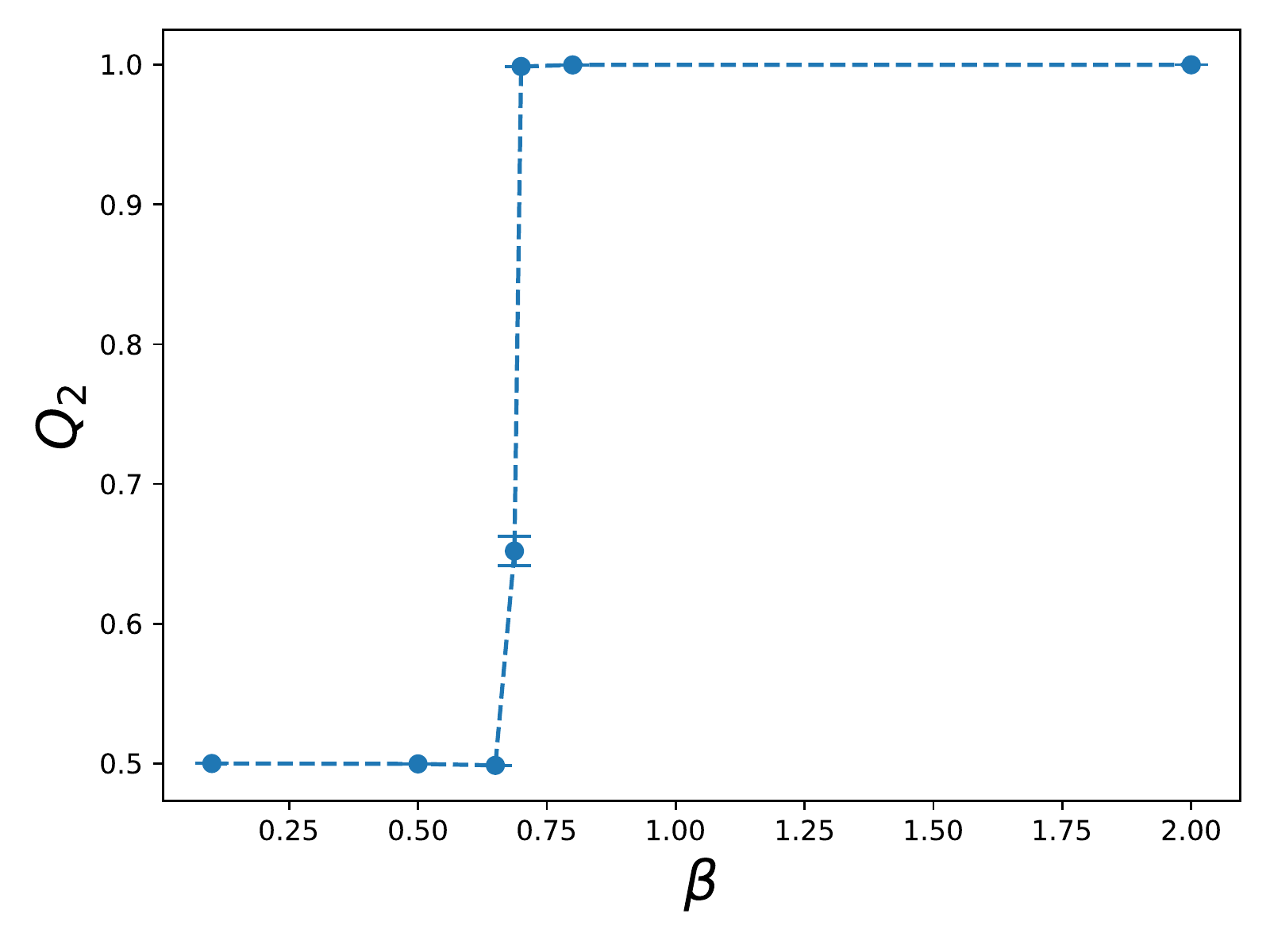}
    }
  \end{center}\vskip-0.7cm
  \caption{$Q_1$ (top panel, obtained on $20^3$ lattices) and $Q_2$ (bottom panel, obtained on $20^3$ lattices) 
    as functions of $\beta$ for the 3D 5-state ferromagnetic Potts model.}
\label{Potts_MC}
\end{figure}

\begin{figure}
  \begin{center}
    
\includegraphics[width=0.4\textwidth]{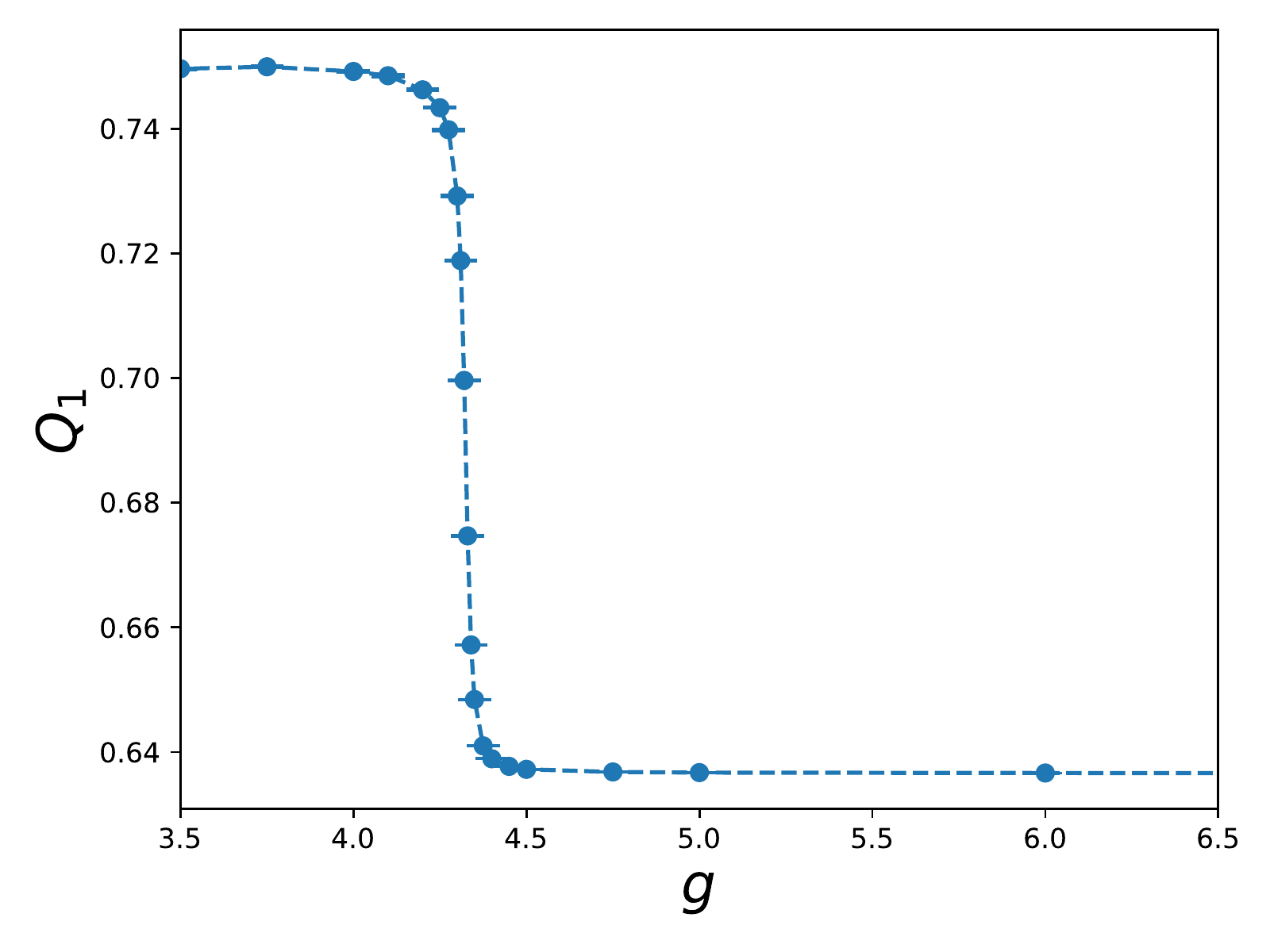}
  \end{center}\vskip-0.7cm
  \caption{$Q_1$ as a function of $g$ for the 3D plaquette model. The
  results are obtained on $32^3$ cubic lattices.}
\label{plaq_1}
\end{figure}

\vskip0.5cm
{\bf The construction of the configurations used for the NN testing} ---
As pointed out in the main text, bulk quantities satisfying certain conditions are suitable for
employing to construct the configurations used for the NN testing. 
For all the three considered models, we find that the first and the second Binder ratios ($Q_1$
and $Q_2$) are the appropriate observables that suit these criterions. Indeed, as can be see
in figs.~\ref{O3_MC}, \ref{Potts_MC}, and \ref{plaq_1}, $Q_1$ and $Q_2$ of the three studied systems fulfill
the requirements outlined previously.

We would like to emphasize the fact that when choosing the relevant quantities, like the Binder ratios
$Q_1$ and $Q_2$, only several data points
are sufficient to determine whether the targeted observables meet the required criterion.

The second method of constructing the configurations for the NN prediction are already
detailed in Ref.~\cite{Tan20.2}. Here we will use the $O(3)$ model to show how
to convert the associated spin states into the suitable variables used
for the testing.

For a given $O(3)$ configuration, the associated 1D lattice of 120 sites is built as follows.
First, 120 sites of the given $O(3)$ configuration is chosen randomly and uniformly.
Second, $\psi\,\, \text{mod}\,\, \pi$ for these 120 chosen sites from the given $O(3)$ configuration,
which is ether 1 or 0, are used as the variables for the 120 sites 1D lattice.

\begin{figure}
  \begin{center}
    \vbox{
\includegraphics[width=0.35\textwidth]{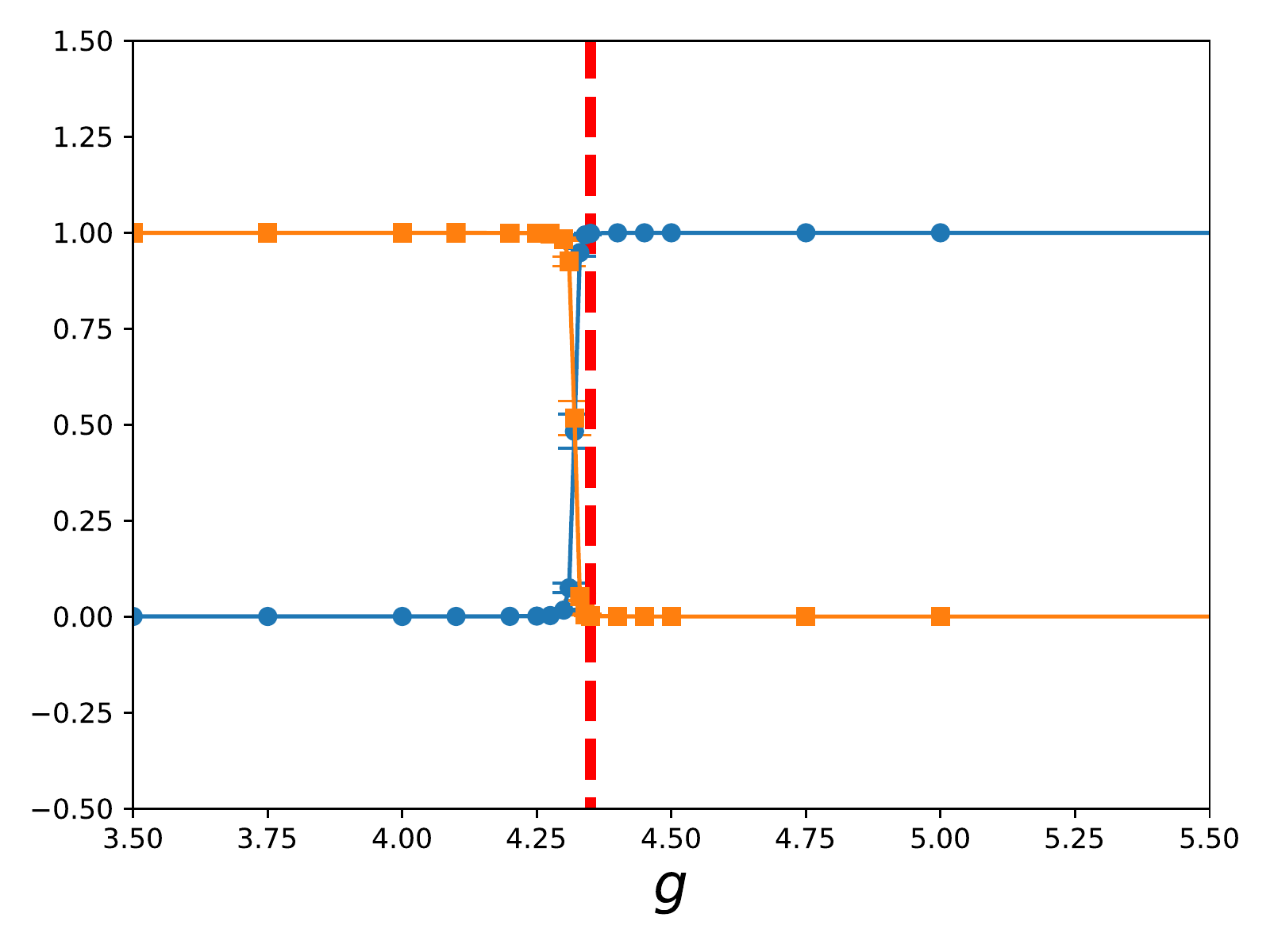}
\includegraphics[width=0.35\textwidth]{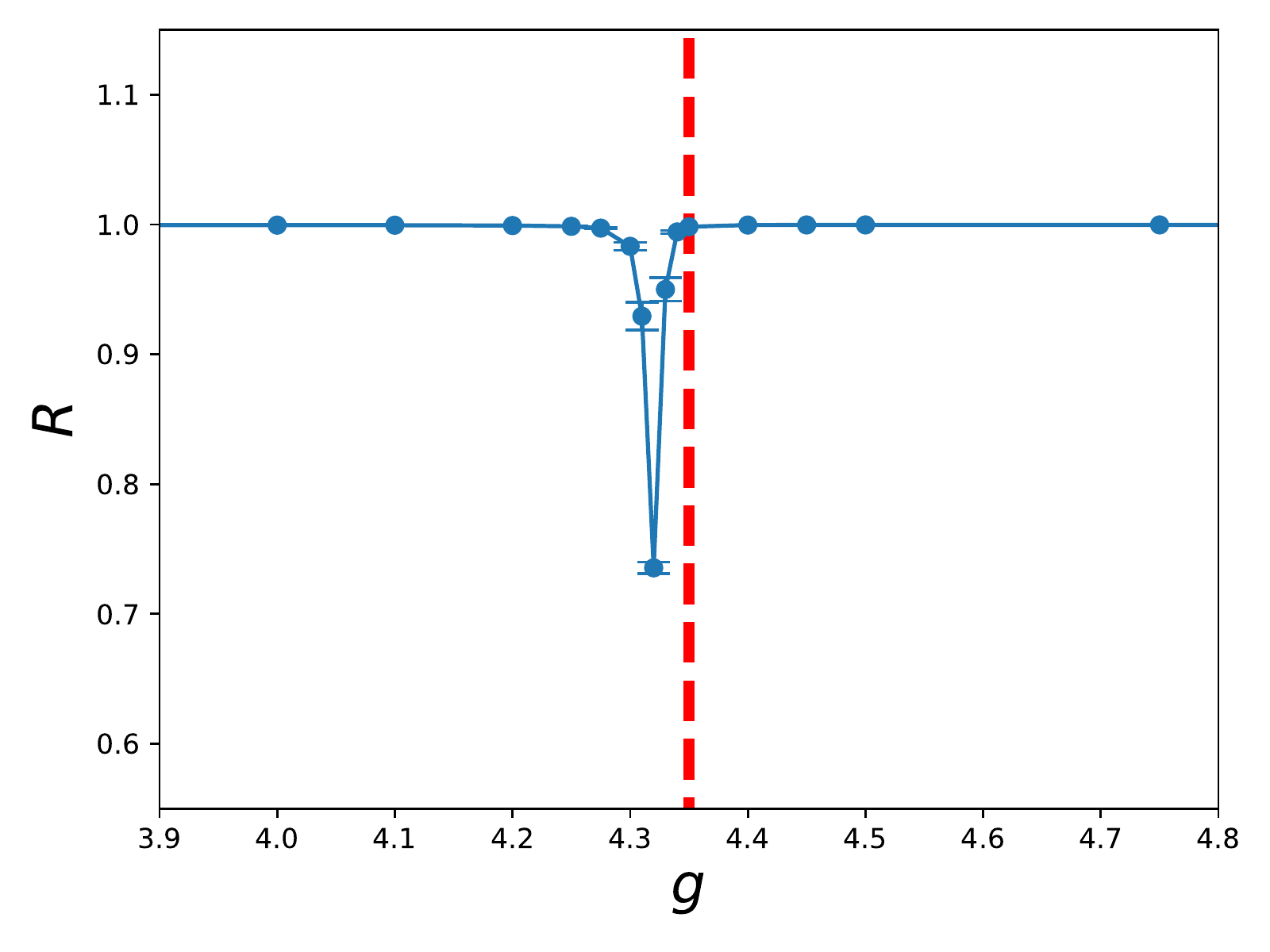}
    }
  \end{center}\vskip-0.7cm
  \caption{Individual component (top) and magnitude $R$ (bottom) of the output vectors as functions of $\beta$ for
    the 3D plaquette model.
    The results are obtained by the NN trained on a 2D square lattice with linear size $L=16$, and
	the configurations employed for the prediction are based on $Q_2$ (on $32^3$ lattices)
    of the considered model. The vertical dashed line in both panel is the expected (lower bound of) $g_c$.}
\label{plaq_2}
\end{figure}

\vskip0.5cm
{\bf More NN results} ---
In the main text, the NN trained on a 1D lattice of 120 sites is used and here we present more outcomes obtained using this NN.
Moreover, we also train two NNs on a 1D lattice of 200 sites and on a 16 by 16
square lattice. The corresponding results will be demonstrated as well.
These additional results again confirm the effectiveness of our extremely efficient
NN approach for investigating the critical phenomena.

Fig.~\ref{plaq_1} shows the quantity $Q_1$ of the 3D plaquette model. The results in the figure
indicate the associated critical point $g_c$ is between $g=4.0$ and $g=4.5$.
In addition, the outcomes of the NN (trained on a 16 by 16 square lattice) using the crossing and the magnitude methods are
demonstrated as the top and the bottom panels of fig.~\ref{plaq_2}. The dashed
vertical line in both panels is the lower bound of $g_c$ estimated using the relevant data.
Clearly, the agreement between the NN and the expected results is remarkable.

\begin{figure}
  \begin{center}
    \vbox{
\includegraphics[width=0.35\textwidth]{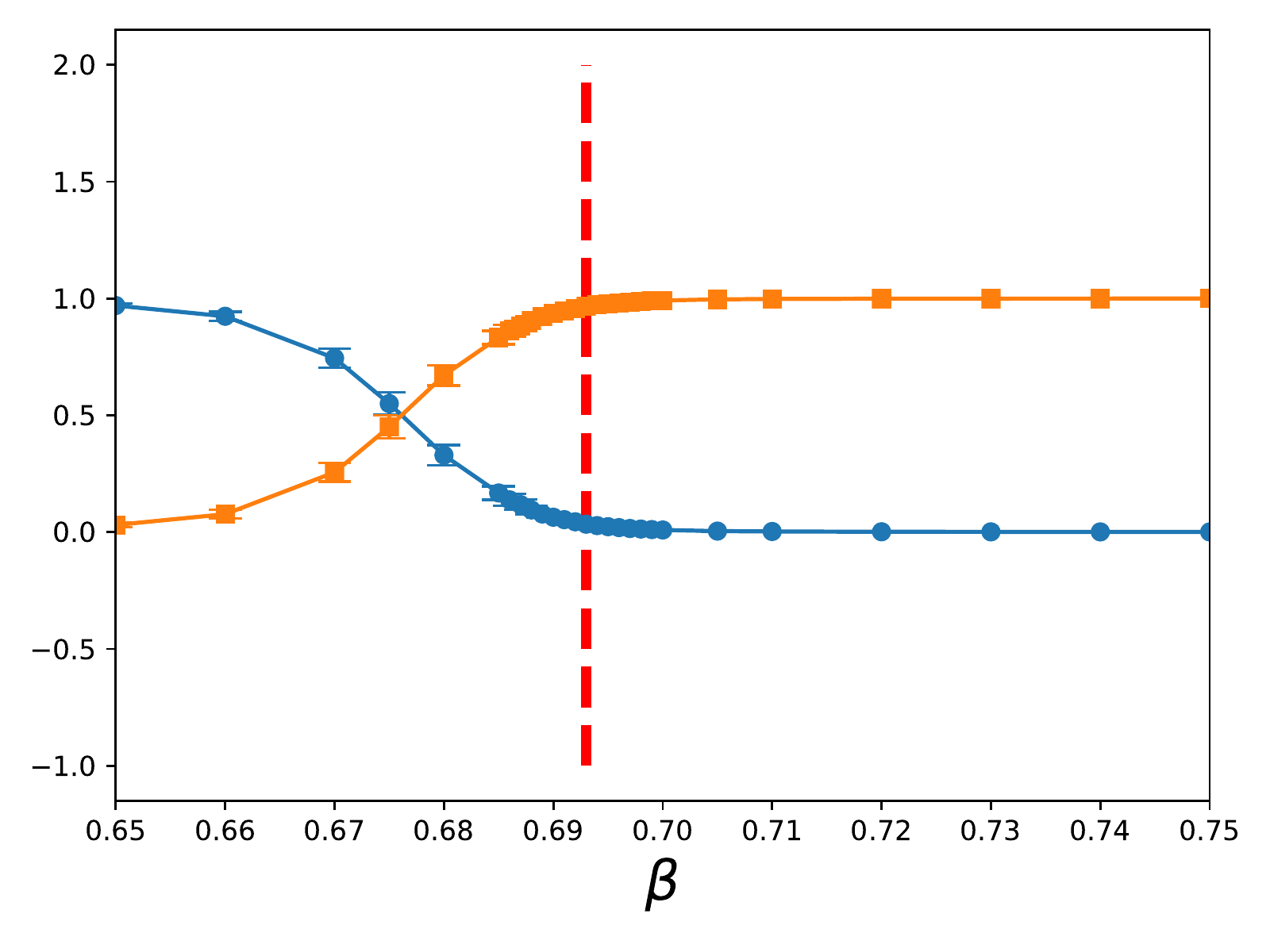}
\includegraphics[width=0.35\textwidth]{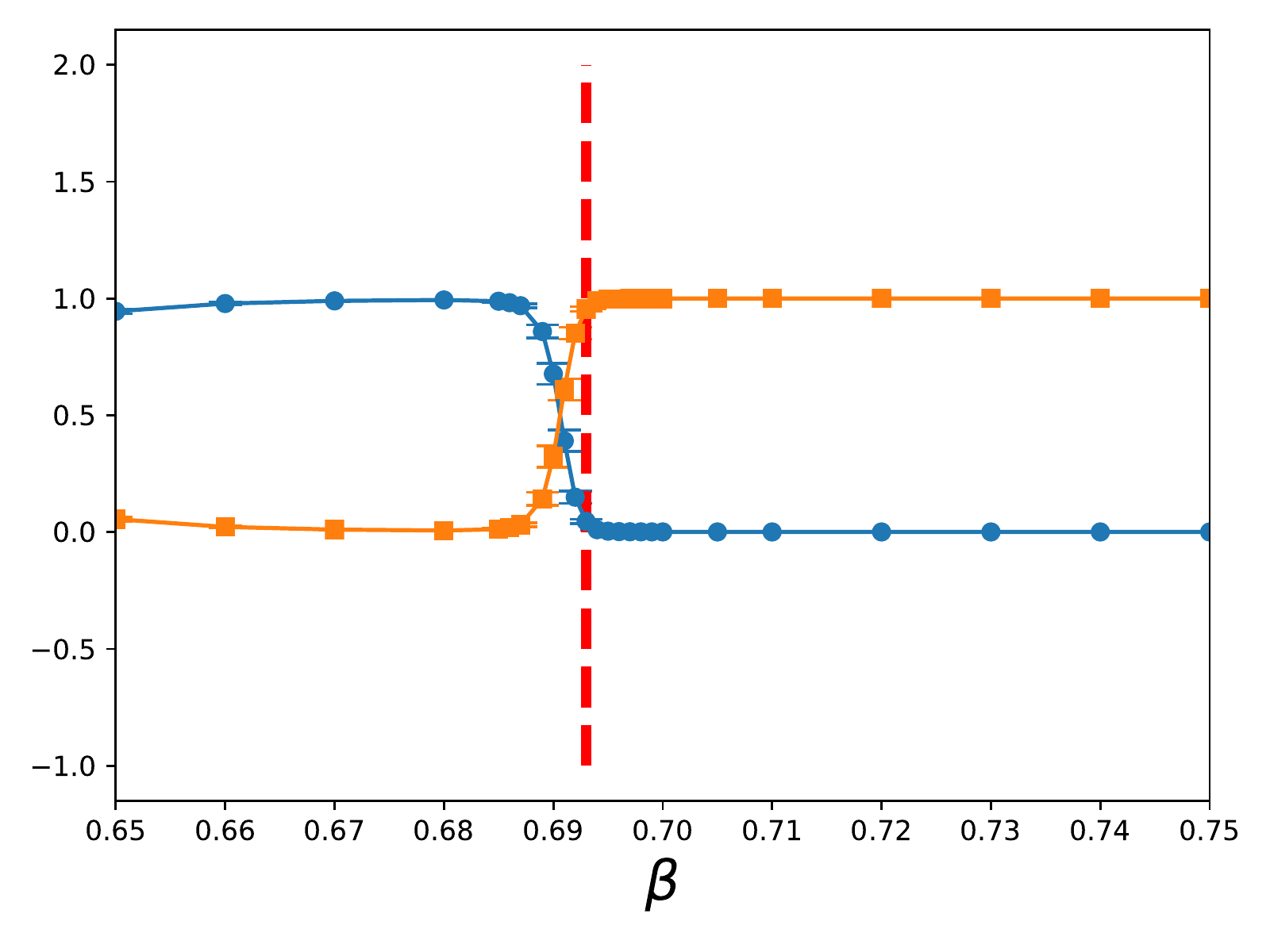}
    }
  \end{center}\vskip-0.7cm
  \caption{Individual component of the output vectors as functions of $\beta$ for
    the 3D classical $O(3)$ model. The dashed line is the expected $\beta_c$. 
    The results are obtained by the NN trained on a 1D lattice of 120 sites, and
    the configurations employed for the prediction are based on $Q_1$
    of the considered model.}
   
\label{more_O3_1}
\end{figure}

Fig.~\ref{more_O3_1} shows the crossing associated with the 3D classical $O(3)$
model. The outcomes are obtained using a NN trained on a 1D lattice of 120 sites
and the configurations employed for the prediction are based on the bulk quantity $Q_1$
of the studied model. Moreover, the $Q_1$ for calculating the outcomes of the top and the bottom panels
are determined on lattices with linear system sizes
$L=12$ and $L=48$, respectively.
As can been seen from the figure, the crossing points move toward the expected
$\beta_c$ as the linear system size $L$ increases.

\begin{figure}
  \begin{center}
 
\includegraphics[width=0.35\textwidth]{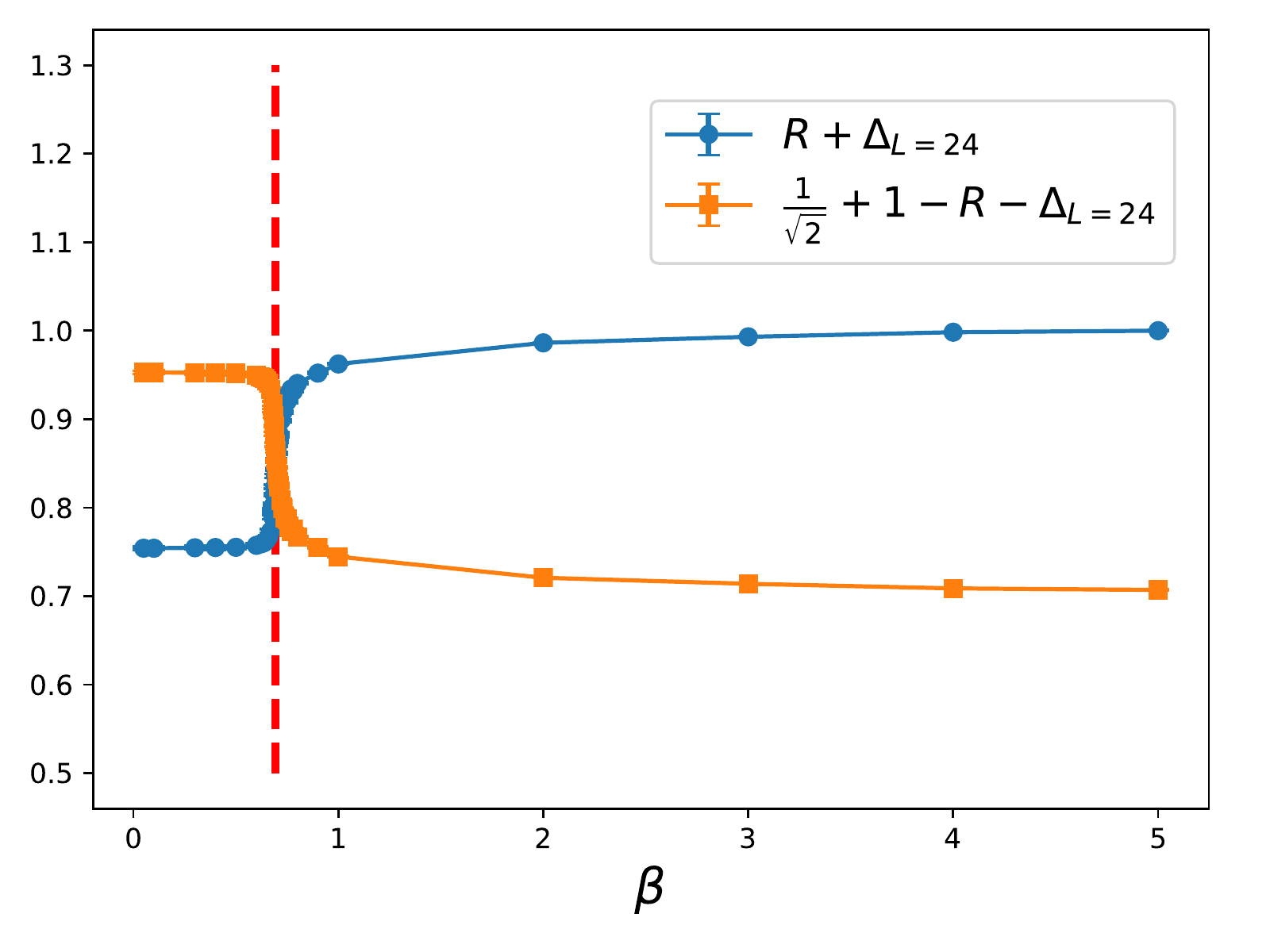}

  \end{center}\vskip-0.7cm
  \caption{$R+\Delta$ and $1/\sqrt{2}+1-R-\Delta$ as functions of $\beta$ for the
    3D classical $O(3)$ model. The dashed line is the expected $\beta_c$. 
    The results are obtained by the NN trained on a 1D lattice of 200 sites, and
    the configurations employed for the prediction are based on the microscopic spin
    states of the considered model.}
\label{more_O3_2}
\end{figure}

The results using configurations based on the detailed spin states of the
$O(3)$ model for the prediction are shown in fig.~\ref{more_O3_2}.
The demonstrated results are obtained with NN trained on a 1D lattice of
200 sites. 

\begin{figure}
  \begin{center}
    \vbox{
\includegraphics[width=0.35\textwidth]{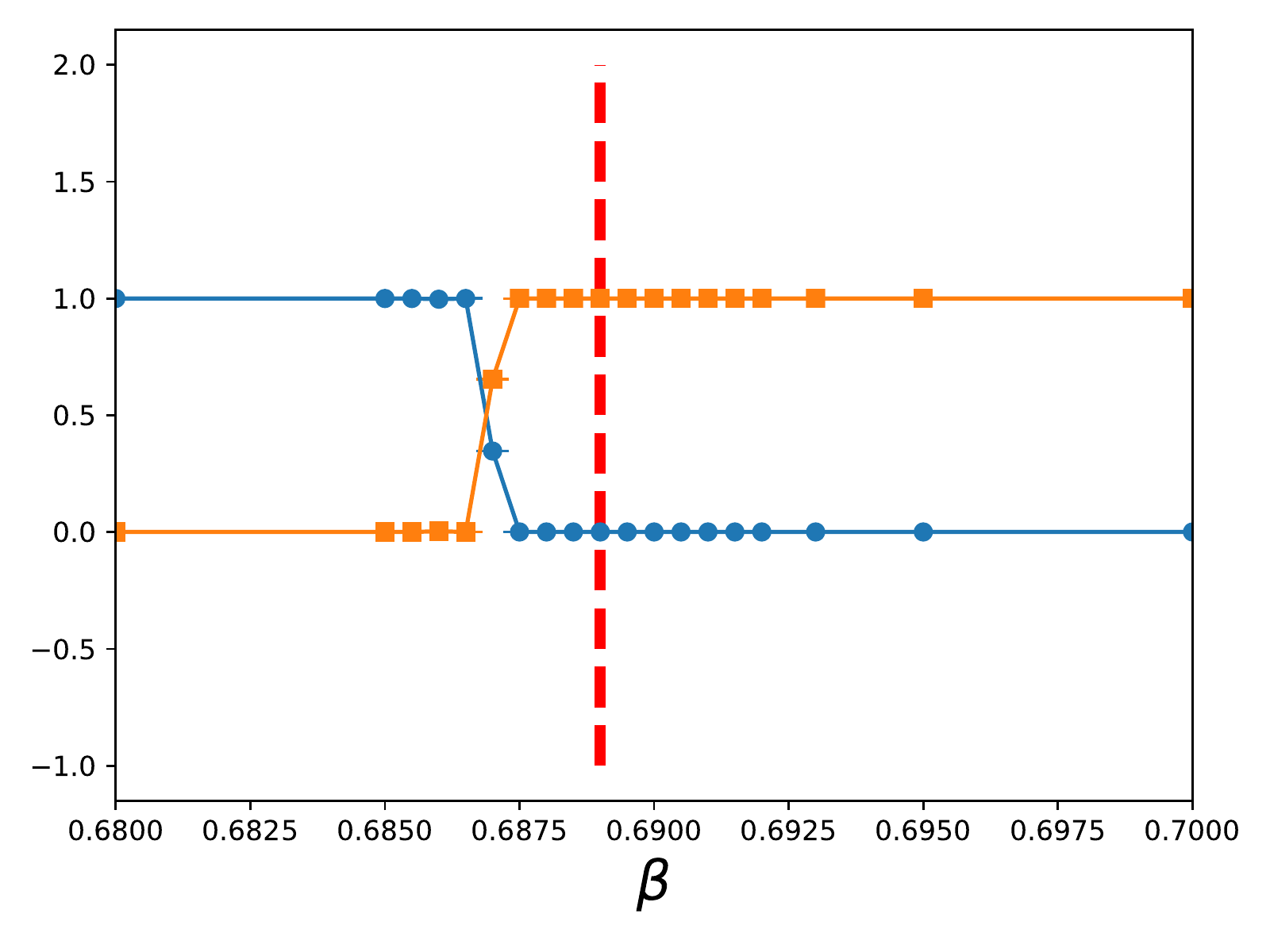}
\includegraphics[width=0.35\textwidth]{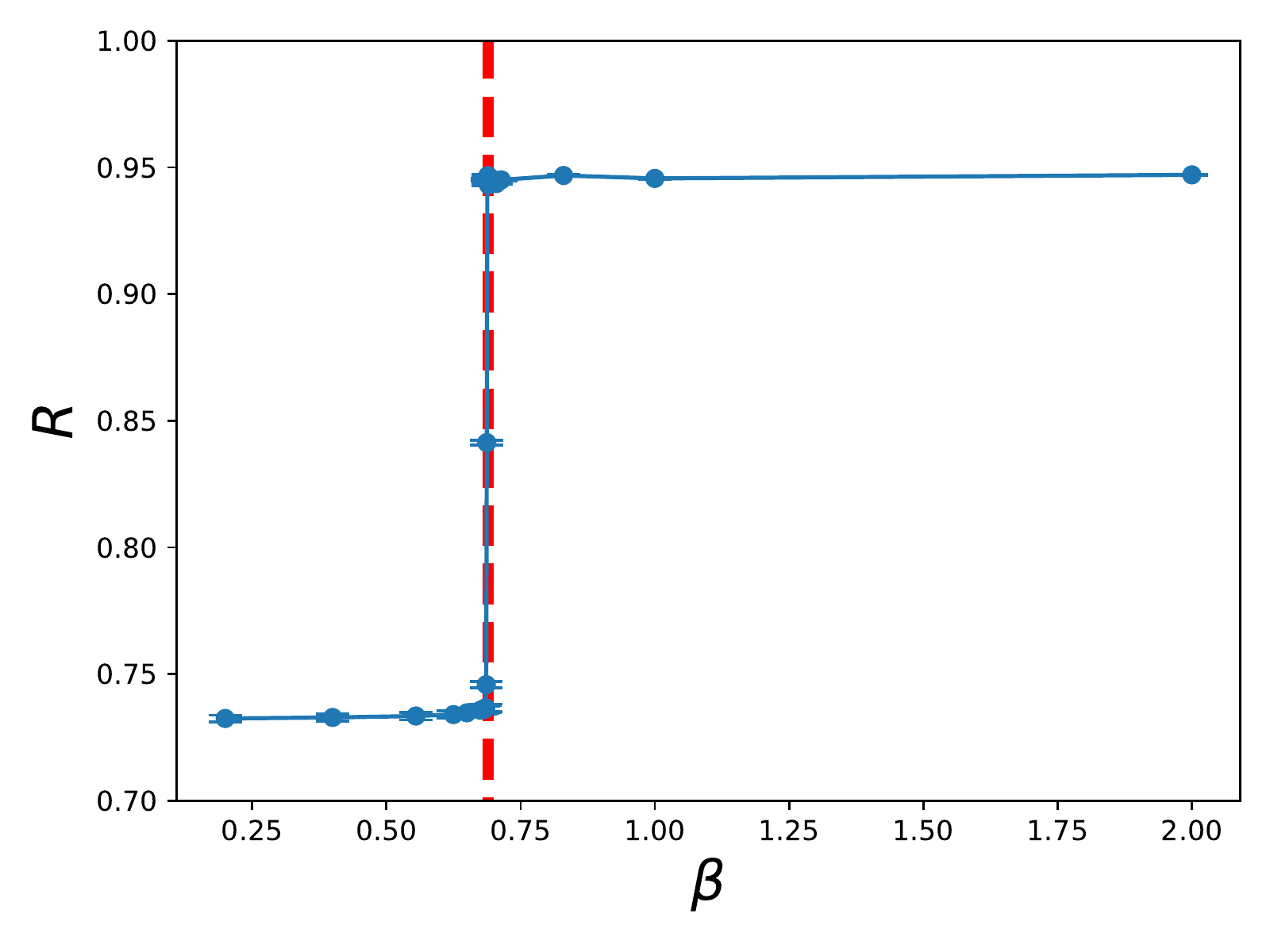}
    }
  \end{center}\vskip-0.7cm
  \caption{(Top panel) Individual component of the output vectors as functions of $\beta$ for
    the 3D 5-state ferromagnetic Potts model.
    The results are obtained by the NN trained on a 1D lattice of 120 sites.
	The configurations employed for the prediction are based on $Q_1$ (on $20^3$ lattices)
    of the considered model.
    (Bottom panel) The magnitude $R$ of the NN outputs as a function of $\beta$ for the 3D 5-state ferromagnetic Potts model.
    The results are obtained by the NN trained on a 1D
    lattice of 200 sites.
    The configurations employed for the prediction are based on the detailed Potts configurations
	(on $20^3$ lattices) of the considered model.}
\label{more_potts_1}
\end{figure}

Similarly, same scenario is observed for the 3D 5-state Potts model, see fig.~\ref{more_potts_1}.
It is interesting that although the targeted phase transition associated with
the 3D 5-state Potts model is first order, the trained NN is capable of estimating
the corresponding $\beta_c$ with high accuracy.

\vskip1.0cm
{\bf Comments and Remarks} ---

In this final part of the supplemental materials, several remarks (comments) regarding
the NN methods proposed in the main text are listed.

\begin{enumerate}
\item{The only requirement needed in our approach is the availability of bulk observables satisfying certain conditions
  mentioned explicitly
  in the main text. It is obvious that such requirements are met in almost all the studies (both theory and experiment)
  associated with phase transitions.}

  \item{The approach demonstrated here does not depend on the microscopic details of considered models. Therefore, it
    is anticipated that our method is applicable for (any) other systems when phase transitions are concerned.}
  \item{Crossings may take place at high temperatures other than the targeted critical points. With the idea
    of using appropriate bulk quantities as input for the NN calculations,
    these situations
    have no influence on our high precision determination of the studied phase transitions.}
  \item{In our NN calculations, only the kernel $L_2$ regularization (with the argument being set to 1)
    is considered. Such a strategy works well for all the studied models.}
  \item{Although further investigation is required, for the quantum plaquette model, using microscopic spin states to produce configurations for the prediction
    does not seem to work. 
  }
  \item{For the 3D 5-state ferromagnetic Potts model, the second method of constructing 1D configurations of 120 sites for the NN prediction
    proceeds as follows.
    First, 120 sites of a given Potts configurations are chosen one by one (randomly and uniformly).
    Second, whenever the Potts variable of the picked site is 1 or 3 (2 or 4), the associated 1D lattice position is assigned the
    integer 0 (1). If the Potts variable is 5, then 1 and 0 are given to the corresponding (1D lattice) site with equal probability.  
  }
  \item{We have carried out a NN calculation to determine the $T_c$ of the 3D $O(3)$ model using the standard procedure. In particular,
    4 temperatures are chosen (two are below and two are above $T_c$) and the corresponding (totally 4000) spin state configurations
    on $48^3$ lattices
    are used for the training. The time taken to complete such a standard training step is around 400 times the time required for
    the NN training (on a 200 sites 1D lattice) conducted in this study. 
  }

\end{enumerate}

\end{document}